\documentclass{jltp} 
\usepackage{graphicx} 
 
\def\be{\begin{equation}} 
\def\ee{\end{equation}} 
 
\title{Temperature Increase of Highly-Polarized Fermi Liquids in Spin-Echo Experiments} 
\author{R. Ragan, K. Grunwald, and C. Glenz} 
\address{Department of Physics, University of Wisconsin at La Crosse\\ La Crosse, WI  54601, USA} 
 \runninghead{Ragan {\it et al.}}{Temperature Increase of Highly-Polarized Fermi Liquids ...} 
 \begin{document} 
 \maketitle 
\begin{abstract} 
We show that there are restrictions on the maximum  
tipping angle that can be used without significantly  
raising the temperature of the $^{3}$He distribution in high B/T spin-echo experiments with pure liquid 
$^{3}$He and $^{3}$He-$^{4}$He solutions. The temperature increase occurs during the diffusion process as quasiparticles in mixed-spin states are scattered and converted into thermal excitations at the spin-up and spin-down Fermi surfaces. This temperature increase can mimic the effects of zero temperature  
attenuation, leading to a higher values of the measured anisotropy  
temperature $T_{a}$. We analyze the dependence of  
the increase on polarization, initial temperature,  
$^{3}$He concentration, and tip angle, and estimate the size  
of the effect in recent experiments. 
\end{abstract} 

\section{INTRODUCTION} 
The question of whether zero-temperature attenuation exists in the transverse spin dynamics  
of spin-polarized normal Fermi fluids continues to attract experimental and theoretical interest. One idea\cite{Meyer94} is that in strongly polarized Fermi fluids, the gap between the spin-up and spin-down Fermi surfaces is available for scattering quasiparticles in mixed-spin states, leading  
to a finite transverse relaxation time $\tau_{\perp} \propto (T^{2}+T_{a}^{2})^{-1}$ as $T \rightarrow 0$, with an   
anisotropy temperature $T_{a}$ that is of the order of the gap energy $\sim E_{Zeeman}=2\beta B$,  where $\beta$ is the magnetic moment and $B$ is the effective external magnetic field. For very dilute mixtures with {\it s}-wave interactions, Jeon and Mullin\cite{GS} found $T_{a}=2\beta B/2  
\pi k_{B} \sim 2-4$ mK with currently available magnetic fields of $8-15$ Tesla. 

 Several high $B/T$ spin-echo experiments have been performed to observe  
zero-temperature attenuation in pure $^{3}$He and dense $^{3}$He-$^{4}$He  
solutions $(x_{3}>1\%)$, which have degenerate temperature regimes that are accessible with dilution refrigeration. Measurements of the anisotropy temperature have thus far yielded
$T_{a}$($8.8$ T, $x_{3}=1\%$)$=8\pm 4 $ mK, $T_{a}$($8.8$T, $3.8\%$)$=13\pm 2 $ mK, and $T_{a}$($8.8$ T, $6.4\%$)=1$9\pm 3 $mK,\cite{Ager95}  $T_{a}$($8$ T, pure)$=16.4 \pm 2.2$ mK,\cite{Wei93} $T_{a}$($11.3$ T, $6.2\%$)$=13\pm 2 $mK\cite{OB00} and $T_{a}$($11.3$ T, pure)$=12 \pm 2$.\cite{Cand00} Although the experimental evidence for zero temperature attenuation is convincing, the measured values of $T_{a}$ seem rather high compared to the dilute limit prediction, and more puzzling, are not proportional to the magnetic field.  
 
On the other hand, Fomin\cite{Fomin97} argues that when the ground state of a  
polarized system is properly defined, the quasi-particles between the Fermi surfaces are not excitations and do not scatter. He  
has derived a spin wave dispersion relation at $T=0$ which contains no damping up to second order in the wave vector.   A  
recent cw experiment\cite{Verm1} to measure the damping of spin-wave modes ($1^{\circ}$ tip) in a saturated 9.4{\%} $^{3}$He-$^{4}$He  
solution at 7 bar showed no evidence of spin diffusion anisotropy, supporting Fomin's prediction that $T_{a}=0$. Thus the existence of zero temperature attenuation is still a somewhat open question.  
  
In hope of clarifying the situation, spin-echo experiments are being performed at the High $B/T$ Facility at NHMFL ($B=15$T and $T\ge2$mK) with very low concentrations to test the dilute limit prediction of Jeon and Mullin. At this high value of $B/T$ the spin rotation parameter is expected to be very large $\mu M>100$. Under these conditions special care must be taken to avoid experimental defects that might mimic or mask zero-temperature attenuation such as joule heating, boundary effects, instabilities, and inhomogeneous demagnetization fields. Below we describe another effect which can mimic zero-temperature attenuation.
 
In a $\theta-\Delta t - \pi$ spin echo experiment the initially uniform equilibrium magnetization is tipped out of thermal equilibrium by a  
$\theta$ pulse. Because of the field gradient, the transverse  
components of the magnetization mix until, at a time $t> T_{2}^{*}$, the magnetization becomes uniform with polarization $P\cos{\theta}$. Note that the system is still out of equilibrium with $B$, which, for times much less than the longitudinal relaxation time $T_{1}$, has no effect on the nonequilibrium dynamics other than an overall Larmor precession. 
 
As the nonuniformities in the magnetization mix, the (non-equilibrium) temperature of the momentum distribution increases.  The temperature increase occurs as the mixed-spin states between the Fermi surfaces of the  
polarized $^{3}$He gas scatter and are converted into thermal excitations near the spin-up and spin-down Fermi surfaces. The thermodynamics of this irreversible process have been analyzed previously in a study of the polarization dependence of the viscosity in $^{3}$He.\cite{Verm2} Note that at these low temperatures ($T<100$ mK), the specific heat of the background $^{4}$He is negligible compared to the $^{3}$He gas.\cite{And}
  
A simple example illustrates the effect. Consider two $T=0$  Fermi distributions, one unpolarized and one completely polarized.  The polarized distribution has a Fermi momentum  that is $2^{1/3}$  
times greater than the unpolarized distribution, and a total  kinetic  
energy which is $2^{2/3}$ times greater. If the polarized distribution is tipped by  $90^{\circ}$ and allowed to relax by diffusion in a field gradient, its temperature must increase from  $T=0$ to $T_{final}=(2^{2/3}-1)T_{F}=0.4438 T_{F}$ since the total kinetic energy is conserved. For a small tip angles, the effect is smaller since the effect of the spin-echo experiment on the polarization is $P \rightarrow P \cos{\theta}$. 

In the following, we calculate the temperature increase for various tipping angles, initial temperatures and concentrations, with special attention paid to the NHMFL experiments. We also estimate the size of the effect in past measurements of the anisotropy temperature. 

\section{FERMI DISTRIBUTION CALCULATIONS}
Consider a polarized ideal Fermi gas in thermal equilibrium with a magnetic field $B$. So that the following applies to the NHMFL experiments, we shall assume $B=15$ T, $T=2$ mK, and $x_{3}=0.2\%$ ($T_{F}=40.6$ mK).  The explicit formulas for the  normalized particle densities are, 

\begin{figure} 
\centerline{\includegraphics[height=2.5in]{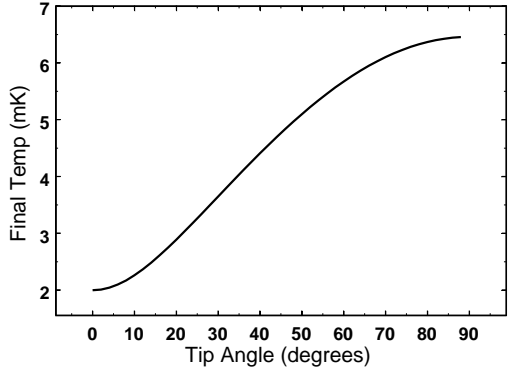}} 
\caption{Final temperature as a function of tip angle for $T=2$ mK, $B=15$ T, and $x_{3}= 0.2 \%$. } 
 \label{fig:angle} 
 \end{figure} 
\be 
n_{\uparrow \downarrow}= \frac {3}{2} 
\int_{0}^{\infty}\frac{x^{2}dx}{1+e^{(-\tilde{\mu}+x^{2}\mp b)/y}} 
\label{dist} 
\ee 
\begin{figure} 
\centerline{\includegraphics[height=2.5in]{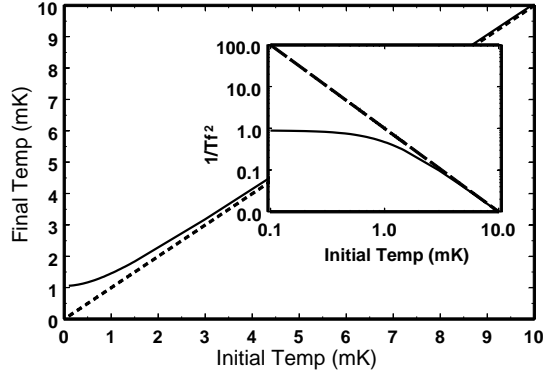}} 
 %
\caption{Final temperature as a  
function of the initial  
temperature for a $10^{\circ}$ tip angle. The dotted line is the initial  
temperature.} 
 \label{fig:temp} 
 \end{figure}
where $y=T/T_{F}=0.0493$, $x=p/p_{F}$, $\tilde{\mu}=\mu/k_{B}T_{F}$, and $b=\beta B/k_{B}T_{F}=0.615$. Solving for the chemical potential numerically, we find  $\tilde{\mu}=0.979$ and a polarization $P= n_{\uparrow}-n_{\downarrow}=0.402$. The kinetic energy of the $^{3}$He distribution is calculated from 

\be 
\epsilon \equiv KE/KE_{F}= \frac {5}{2}  
\int_{0}^{\infty}\left(\frac{1}{1+e^{(-\tilde{\mu}+x^{2}+b)/y}}+\frac{1}{1+e^{(-\tilde{\mu}+x^{2}-b)/y}}\right)x^{4}dx 
\label{energy} 
\ee  
During a spin-echo experiment the system is out of equilibrium with the magnetic field, we may remove the Zeeman energy term in the exponentials and introduce two nonequilibrium chemical potentials. In order to calculate the temperature increase, we re-solve Eqs.\ref{dist}-\ref{energy} for $y$ and $\tilde{\mu}_{\pm}$ with the initial kinetic energy but with $P \rightarrow P\cos{\theta}$. The resulting temperature increase is shown in Fig.1 as a function of the tip angle. For small tip angles the effect is small: for tip angles of $5^{\circ}-10^{\circ}$ the temperature increases from an initial temperature of 2 mk to final temperatures of $2.07-2.26$ mK. However, the final temperature increases rapidly with increasing tip angle to a maximum of 6.5 mK for a $90^{\circ}$ tip. 

In Fig.2 the final temperature is plotted as a function of the initial temperature for a $10^{\circ}$ tip angle. Since the diffusion  
coefficient and $\mu M$ vary like $1/T^{2}$ without zero-temperature  
attenuation, we also plot $1/T_{final}^{2}$ in the inset.
One can see that the temperature increase could mimic anisotropy effects, with an apparent anisotropy temperature of 0.75 mK. This, however, is much lower than the expected anisotropy temperature of 3.8 mK for 15 T. 

For a given temperature and field strength the effect in $^{3}$He-$^{4}$He solutions is nearly independent of concentration (the calculations are not presented here). On the other hand, for pure $^{3}$He the effect is larger because the polarization is enhanced by the factor $(1+F_{0}^{a})^{-1}=3.3$. In previous spin-echo experiments with pure $^{3}$He and magnetic fields of 8 T\cite{Wei93} and 11.3 T \cite{Cand00}, the temperature increases were from initial temperatures of 4 mK and 2.5 mK to final temperatures of 4.57 mK, and 3.5 mK, repectively, which are both consistent with an apparent anisotropy temperature of $\approx$ 2 mK. In both cases the temperature increase was not large enough to mimic the observed anisotropy temperatures of 13-16 mK, but the increase certainly shifted the observed $T_{a}$ upward. For previous experiments in $^{3}$He-$^{4}$He solutions the effect is hardly noticeable. 

\section{DISCUSSION: EFFECT ON SPIN ECHO EXPERIMENTS}
Next, we consider the effect that the temperature increase has on the spin- echo data and the resulting measurements of $D_{\perp}$ and $\mu M$.

\begin{figure} 
 \centerline{\includegraphics[height=2.5in]{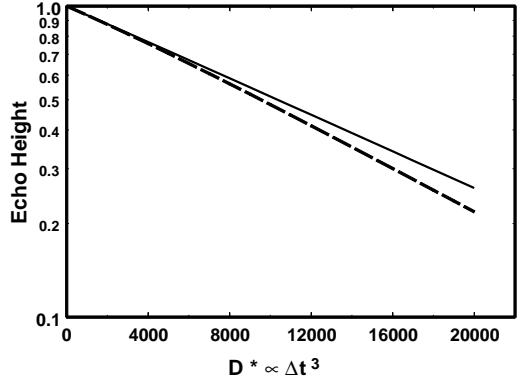}} 
 %
 \caption{Calculated spin echo heights vs. pulse interval for $T_{a}=0$. The dotted line results when the temperature increase is included. } 
 \label{fig:echo} 
 \end{figure} 
If the time to restore thermal equilibrium in the momentum distribution $\tau_{\perp}$ is small compared to the diffusion time $(D_{\perp}k^{2})^{-1}$, then in computer simulations the distribution can treated as having an instantaneous temperature $T[P(t)]$ during the spin-echo experiment. When the temperature increase is small compared to the anisotropy temperature $T_{a}$, the effect on the spin echo data is barely noticeable. However, if Fomin is correct and $T_{a}=0$ the temperature increase can mimic the effects of zero temperature attenuation. To illustrate this, Fig.3 shows the effect of the temperature increase on the calculated spin echo heights for $\theta=10^{\circ}$, $\mu M =100$ and $T_{initial}=2$ mK as in the NHMFL experiment, with $T_{final}=2.26$ mK and $T_{a}=0$.  
Since the temperature increase develops after a time $T_{2}^{*}$ the initial slope of the $\log{h}$ vs. $t^{3}$ is unaffected. Within typical experimental error bars the echo height curve still appears linear, but with an somewhat increased slope. The slope, which is proportional to $D_{\perp}/(\mu M)^{2}$, is increased since the temperature increase reduces both $D_{\perp}$ and  $\mu M$ by the same factor.  By similar arguments, the phase of the echo (not shown) is unaffected. When the echo height and phase data are fitted to lines, the resulting values of $D_{\perp}$ and $\mu M$ are reduced by 13{\%}, which is consistent with the (false) anisotropy temperature of $T_{a}\approx 0.7$ mK shown in Fig.2. 

From these calculations of the temperature increase, we can eliminate this effect as the explanation for the observation of anisotropy temperatures in previous spin-echo experiments, although this effect will become increasingly important at higher values of $B/T$.

\section*{ACKNOWLEDGEMENTS}
This research was supported by NSF grant DMR-0071706 and an NSF-REU Supplement.

\end{document}